\documentclass[12pt,a4paper,JCP]{article}
\usepackage[textheight=23cm,textwidth=16cm]{geometry}
\usepackage{amsmath}
\usepackage{graphicx}
\usepackage{threeparttable}
\usepackage{multirow}
\usepackage{cite}
\usepackage{color}
\usepackage{ulem}
\usepackage[american]{babel}

\begin{document}

\begin{center}

{\LARGE\bf
 Density-Matrix Renormalization Group Algorithm with Multi-Level Active Space
}

\vspace{2cm}

{\large
Yingjin Ma\footnote{E-Mail: yingjin.ma@yahoo.com. Present address: ETH Zurich, Laboratorium f\"ur Physikalische Chemie,
Vladimir-Prelog-Weg~2, 8093 Zurich, Switzerland}, Jing Wen, Haibo Ma\footnote{E-Mail: haibo@nju.edu.cn.}
}\\[4ex]

 Key Laboratory of Mesoscopic Chemistry of MOE, \\
School of Chemistry and Chemical Engineering, \\
Institute of Theoretical and Computational Chemistry, \\
Nanjing University, Nanjing 210093, China

\vspace{5cm}

\vfil

\end{center}

\begin{tabbing}
Date:   \quad \= \today \\

\end{tabbing}

\newpage

\begin{abstract}

The density-matrix renormalization group (DMRG) method, which can deal with a large active space composed of tens of orbitals, is nowadays widely used as an efficient addition to traditional complete active space (CAS)-based approaches.
In this paper, we present the DMRG algorithm with a multi-level (ML) control of the active space based on chemical intuition-based hierarchical orbital ordering, which is called as ML-DMRG with its self-consistent field variant ML-DMRG-SCF.
Ground and excited state calculations of H$_2$O, N$_2$, indole, and Cr$_2$ with comparisons to DMRG references using fixed number of kept states ($M$) illustrate that ML-type DMRG calculations can obtain noticeable efficiency gains. It is also shown that the orbital re-ordering based on hierarchical multiple active subspaces may be beneficial for reducing computational time for not only ML-DMRG calculations but also DMRG ones with fixed $M$ values.

\end{abstract}

\newpage

\section{Introduction}

\quad {One of the main challenges in the current community of theoretical chemistry is the description of the systems with complicated electronic structures, e.g. the transition metal complexes, open shell and excited electronic states, and the bond-breaking/formation \cite{QCbook1, CI_REV, QCbook2, MRCC}. In such cases, several leading components exhibit in the wave function due to quasi-degeneracy and accordingly the accuracy of the Hartree-Fock (HF) approach and those correlation models which rely on the HF assumption are deteriorated.} For dealing with these quasi-degeneracy problems, multi-reference (MR) approaches \cite{CI_REV, MRCC}, which use a number of important reference states instead of only one in the single-reference (SR) approaches as the basis to build further possible excitation configurations, are usually required. In MR approaches, the orbitals
are usually taken from a multi-configurational (MC) self-consistent field
(MCSCF) \cite{MCSCF1, MCSCF2, MCSCF3, MCSCF4} calculations.
In a most popular MC approach, the complete active space (CAS) self-consistent field (CASSCF)\cite{CASSCF1} method, all possible configurations in the active space are constructed, thus the quasi-degeneracy can be taken into account. Thanks to the latest methodological developments \cite{CI_REV, MRCC, CASPT2, NEVPT2_1, NEVPT2_2, Hanauer12, Maitra12, Datta12, Shen13}, great progresses have been made for some of the challenging MR issues such as exploring the multi-radical character of one- and two-dimensional graphene  nanoribbons (GNRs) \cite{Plasser13} and analyzing the magnetic interactions in molecules and highly correlated materials \cite{Malrieu14}. However, the size of the active space used in CASSCF is usually no larger than (16e,16o) due to the exponential growing of the computational cost with the number of active orbitals and electrons \cite{CI_REV, MRCC}, and this greatly hampers the routine applications of CAS-based MR approaches to the large molecular systems.

In recent years, the \textit{ab initio} density-matrix renormalization group (DMRG) method \cite{DMRG_REV, DMRG2, DMRGRCM, TM08, Zgid08, Block, Wouters12, CheMPS2, QCDMRG_REV1, QCDMRG_REV2, Mitrushenkov01, ASDDMRG, Legeza03-1, Moritz05, Wouters14-2, Nakatani13, dmrgct, Cr2_dmrgpt2, dmrgmrci, Sharma14, Rissler06, Legeza03, DBSS, Legeza04, DBSS2, Hachmann07, Mizukami13, DMRGAPP, DMRGAPP2, DMRGCASSCF, DMRGSCF, Ma13, Wouters14, Luo10, QCDMRG_TRAN, Sebastian, Kurashige09, Quan_inf, DMRG_env, DMRG_ctrl, DMRG_qinf, Graph, Kurashige14, Daul00, Chan03, Chan04, Chan04-2, Chan05, Hachmann06, Dorando07, Chan08, Yanai09, Dorando09, Chan09, Nakatani14, Zgid08-2, Moritz05-2, Moritz07, Podewitz11, Boguslawski12-2, Boguslawski12, Tecmer14, Mizukami10, Lan14, Kurashige14-2, Neuscamman10, Yanai12, Wouters13, Szalay15, Yanai15, Knecht14, Lan15, Hedegard15, Olivares-Amaya15, Dresselhaus15} emerges as a promising MC quantum chemical approach which can deal with active spaces larger than those by traditional CAS-based methods. In the DMRG method which originated from the community of condensed matter physics by the pioneering work of White \cite{DMRG}, the $M$ eigenvectors with largest eigenvalues of the reduced density-matrix (RDM) of sub-systems constitute a truncated basis set for the purpose of reducing the freedom of the Hilbert space. Because its computational cost scales only polynomially with the number of active orbitals and the number of retained renormalized basis states $M$
\cite{DMRGRCM}, DMRG is nowadays regarded as an efficient alternative to the full configuration interaction (FCI) or the CAS configuration interaction (CASCI) method which has an exponential scaling.
With the latest improvements of the DMRG method (e.g. spin-adaptation \cite{Zgid08, Block, Wouters12, CheMPS2}, matrix-product state (MPS) \cite{MPSDMRG, MPS1, Ostlund95, Verstraete04} or tensor network state (TNS) \cite{Nakatani13, Szalay15} representation of the wavefunction, and the incorporation of dynamical correlations \cite{dmrgct, Cr2_dmrgpt2, dmrgmrci, Sharma14, Kurashige14, Hedegard15} as well as the consideration of relativistic effect \cite{Moritz05-2, Knecht14, Lan15}), DMRG may be applied to molecular systems with exceedingly large-size CAS which are completely unaffordable by traditional approaches. For example, Hachmann et al. \cite{Hachmann07} firstly illustrated the polyradical nature in higher acenes through (50e, 50o) DMRG calculations, and later Mizukami, et al. \cite{Mizukami13} further extended the radical nature study to GNRs by (84e, 84o) DMRG calculations and demonstrated that the mesoscopic size effect that comes into play in quantum many-body interactions of a large number of $\pi$ electrons is responsible for the polyradical nature in GNRs.
Very recently Kurashige et al. \cite{DMRGAPP} presented the first \textit{ab initio} evidence of possible non-adiabatic chemical pathways of Mn$_4$CaO$_5$ cluster in photosystem II by virtue of (44e, 35o) DMRG calculations,
and Sharma et al. \cite{DMRGAPP2} critically examined the validity of the consensus phenomenological models of iron-sulfur clusters with the help of (30e, 32o) and (54e, 36o) DMRG benchmark calculations and found the widely used Heisenberg double exchange model underestimates the number of states by one to two orders of magnitude.
Nevertheless, it is well-known that in many cases the optimal active orbitals cannot be accurately pre-identified in a simple way and accordingly the optimization of active orbital is usually required for MC and MR calculations. Therefore, it is of crucial importance to perform DMRG calculations with optimized orbitals, which may be self-optimized by energy minimization, along the same lines as CASSCF \cite{DMRGCASSCF, DMRGSCF}. Alternatively, one could also carefully choose various molecular orbitals from other low-level electron-correlation calculations as DMRG basis \cite{Ma13, Wouters14, Luo10}.

An important feature of DMRG is that it allows to capture all types of electron
correlation effects (dynamical and non-dynamical) in a given active space in a consistent way.
In many cases, choosing a fixed number $M$ of preserved states in DMRG-based calculations is feasible but not economic.
Legeza et al. \cite{DBSS, Legeza04, DBSS2} suggested a dynamical block state selection (DBSS) protocol based on a fixed truncation error $\rho_{tr}$ of the subsystem's reduced density matrix instead of using a fixed number $M$ of preserved states in DMRG sweeps. This approach leads to substantial gains in computational efficiency, because such a protocol can automatically adapt the computational demand of a DMRG step at a specified orbital to its special electron-correlation environment.
Since not every renormalized state contributes the same amount to the final energy at every point in the sweep, a fixed constant \textit{M} as used in standard calculations can be a bit inefficient when dealing with very weakly occupied/unoccupied orbitals. Similarly, the truncation by density-matrix threshold similarly is inefficient because a fixed density-matrix weight does not mean the same thing energetically at the ends of the sweep as at the middle.
It is noticeable that the idea of ``multi-level (ML) control of the active space" is popular in the MC approaches \cite{RASSCF1, RASSCF2, CCASSCF, QCASSCF, ORMASSCF, SplitGAS, GASSCF_TIMO, GASSCF}, such as restricted active space self-consistent field (RASSCF) \cite{RASSCF1, RASSCF2} and generalized active space (GAS)-CI/SCF \cite{SplitGAS, GASSCF_TIMO, GASSCF}. In such kind of methods, the whole active space is divided into several subspaces according to their distinct electron-correlation environments, and each subspace is controlled at a different computational level.
For example, GASSCF calculations on the Gd$_2$, and oxoMn(salen) complex by Ma et al. \cite{GASSCF} showed that GAS wave functions achieve the same accuracy as CAS wave functions on systems that would be prohibitive at the CAS level. However, whether the ML concept can be effectively applied to the DMRG method is still unknown.

In this paper, we intend to investigate the possibility and the performance of DMRG calculations using hierarchical orbital reordering of multiple active levels based on chemical intuitions, as commonly used in RAS calculations.
The structure of this paper is as follows. In Sec. II, we present a short description of the DMRG and ML-DMRG algorithms as well as how to combine ML-DMRG with an orbital-optimization scheme. Illustrative numerical examples of the ground and excited states of H$_2$O, N$_2$, Cr$_2$ and indole with comparisons to standard DMRG references are presented in Sec. IV. Finally, we draw our conclusions in Sec.V.

\section{Methodology}

\quad In the MPS ansatz, the DMRG wave function can be expressed as \cite{MPSDMRG}
\begin{equation}\label{MPS}
    |\psi\rangle=\sum_{\sigma_1...\sigma_L}A^1[\sigma_1]A^2[\sigma_2]... A^L[\sigma_L]|\sigma_1...\sigma_L\rangle,
\end{equation}
where matrix $A^i[\sigma_i]$ is for site $i$ with the local state label $[\sigma_i]$, the dimension of $A^i[\sigma_i]$ is $M \times M$ at most, and $L$ is the total number of sites.
The determination of an MPS wave function is a non-linear problem. However, it can be reduced to a linear problem by the sweep algorithm, by which the variational parameters in the MPS wave function can be variationally determined by just iteratively optimizing one or two of the matrices $A^i$ while keeping all other matrices fixed \cite{DMRG}. The MPS wave function can provide both qualitative and quantitative descriptions of the electronic structure, achieved by using small and large $M$. The polynomial scaling (${M^2L^4}+M^3L^3$) of the computational effort (in sweeps) with the size of system is owned by DMRG algorithm \cite{DMRGRCM}.
Usually the \textit{M} is fixed by a specific number (or dynamically adjusted in DBSS \cite{DBSS}).
When applying the DMRG algorithm into quantum chemistry, one need to use it to deal with the second quantized Hamiltonian \cite{QCDMRG_REV1, QCDMRG_REV2},
\begin{equation}\label{SQ}
   \widehat{H}=\sum_{ij}h_{ij}a^+_ia_j+\frac1{2}\sum_{ijkl}V_{ijkl}a^+_ia^+_ja_ka_l
\end{equation}
where $a^+$ and $a$ represent creation and annihilation operators, \textit{i}, \textit{j}, \textit{k}, \textit{l} denote orbitals (e.g. HF canonical molecular orbitals) which are more appreciated than ``sites" by chemists, \textit{h} and \textit{V} are the 1- and 2-electron integrals, respectively.

\begin{figure}[!htp]
 \begin{center}
   \includegraphics[scale=0.30,bb=0 50 1300 1000]{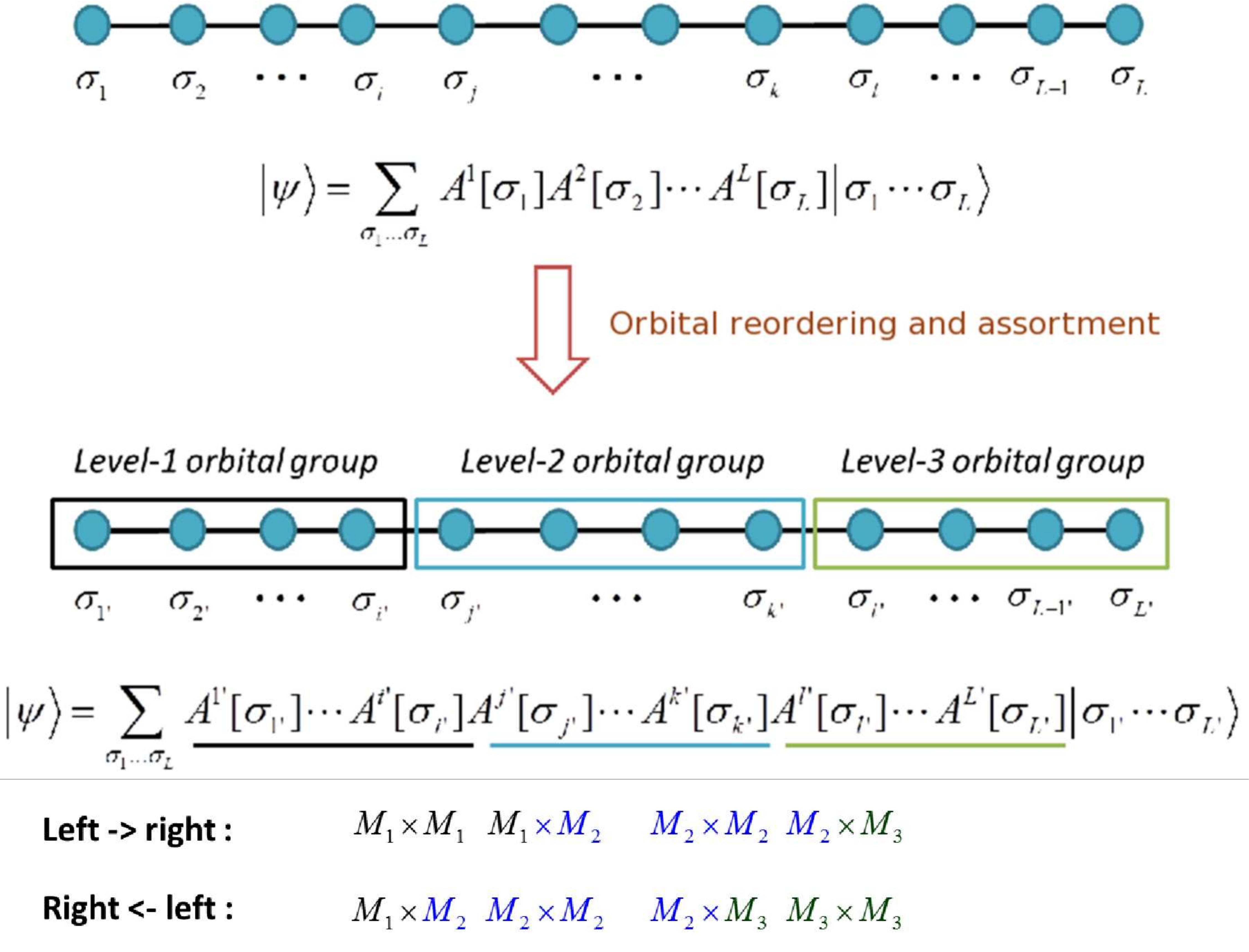}
 \end{center}
 \caption{Illustration of ML-DMRG treatment. The dimension of \textit{A} matrix of the boundary sites(i.e. orbitals) in different sweeping directions are shown at the bottom.}
\label{Fig.MLDMRG}
\end{figure}

In our ML-DMRG framework, we further regroup orbitals into different orbital spaces, and construct the DMRG wave function for the active space with a multiple $M$ treatment as illustrated in Fig.~\ref{Fig.MLDMRG}.
Usually the subspace, which is assigned in the middle of subspaces, should contain the most important orbitals for the purpose of maximizing the efficiency of the sweep algorithm in DMRG \cite{Legeza03}.

\begin{figure}[!htp]
 \begin{center}
   \includegraphics[scale=0.25,bb=0 100 1300 950]{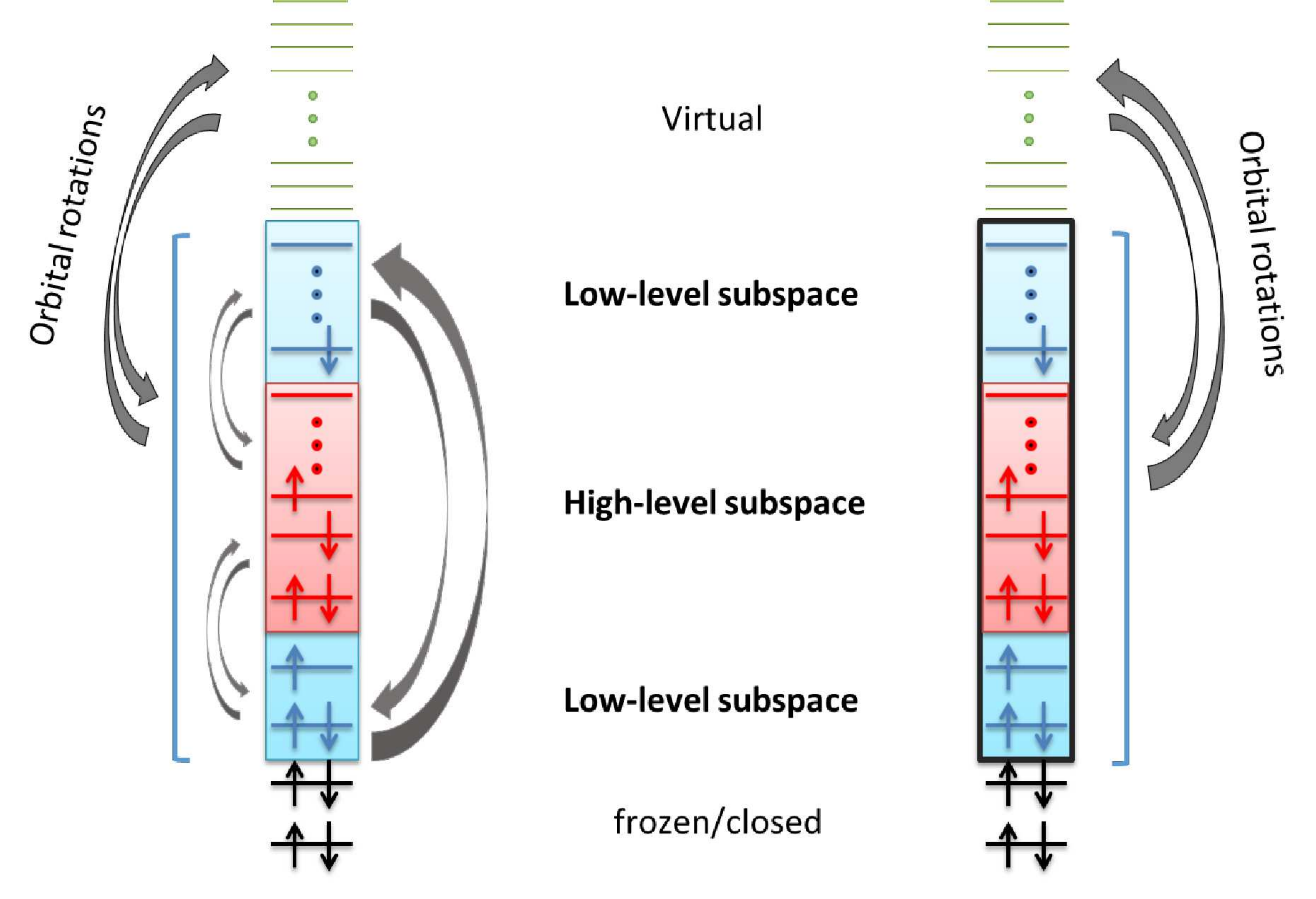}
 \end{center}
  \caption{Treatment of orbital rotations in ML-DMRG-SCF method: 1) Original treatment (left) shows that all the necessary orbital rotations are allowed; 2) Simplified treatment (right) shows that orbital rotations between different subspaces are treated as redundance.}
\label{Fig.ML-DMRG-SCF}
\end{figure}

When combining ML-DMRG with orbital-optimization, a two-step procedure is implemented.
At each iteration, an ML-DMRG calculation is performed first and a reference state is obtained. Then it is followed by orbital-optimization process, according to the super configuration interaction (super-CI) scheme \cite{SuperCI1}. With the new set of orbitals, a new ML-DMRG iteration is performed, until there are no orbital rotations, which indicates that the ML-DMRG-SCF convergence has been reached.
The flowchart of ML-DMRG-SCF is also illustrated in Fig.~\ref{Fig.ML-DMRG-SCF}, and two types of orbital-rotation treatments are used: one is the original treatment, in which all the necessary orbital rotations are allowed; the other is the simplified treatment, in which the orbital rotations between different active subspaces are treated as redundant.

\section{Computational Details}

\quad We use
a locally modified version of the {\sc Block} program \cite{Block} to implement the ML-DMRG calculation. When running the ML-DMRG-SCF calculations, we use a development version of the {\sc Molcas} package \cite{Molcas1}, which contains a {\sc Molcas}-DMRG interface. This {\sc Molcas}-DMRG interface supports {\sc Block} and {\sc Maquis} \cite{Sebastian} QC-DMRG packages. The single point calculations by DMRG-CASCI/SCF, ML-DMRG(-SCF), and geometry optimizations
can be implemented using this interface.
The CASSCF, RASSCF calculations of PEC of N$_2$ are implemented by {\sc Molpro.2010} \cite{Molpro}, and all the other calculations are implemented by the development version of {\sc Molcas} package.

\section{Results and Discussions}

\subsection{Accuracy of ML-DMRG}

\begin{figure}[!htp]
 \begin{center}
   \includegraphics[scale=0.34, bb=0 50 1100 800]{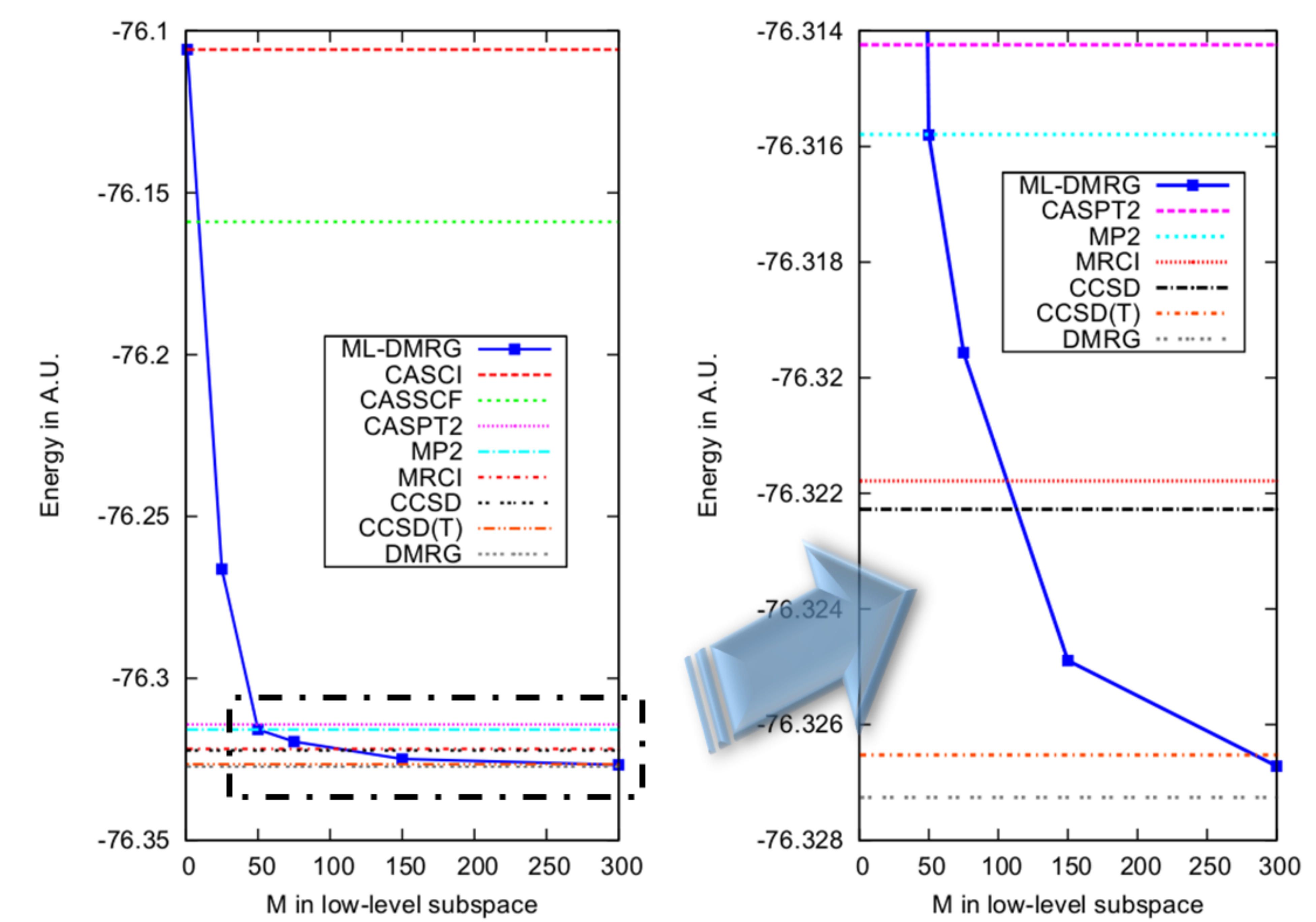}
 \end{center}
  \caption{Results of accuracy analysis along with increased number of preserved states \textit{M} in the low-level subspace of the H$_2$O molecule. Here 4$a_1$ 2$b_1$ 0$a_2$ 1$b_2$ orbitals are chosen as the high-level orbitals with $M$=3000, 7$a_1$ 5$b_1$ 2$a_2$ 3$b_2$ orbitals are chosen as the low-level orbitals with different \textit{M} (shown in figure). The ANO-RCC-VDZP basis is used here.}
\label{Fig.H2o_MLDMRG}
\end{figure}

Firstly, we investigate the numerical accuracy of the ML-DMRG on the water molecule with a ANO-RCC-VDZP basis.
In this example, the core orbitals and the valence orbitals are chosen as the high-level orbitals (10e, 7o) and the remaining orbitals are chosen as the low-level orbitals (17o). In the ML-DMRG calculations, the number of preserved state (\textit{M}) is fixed to 3000 when sweeping among high-level orbitals and the \textit{M} value is range from 1 to 300 when sweeping among low-level orbitals. For comparisons, we also run the CASCI, CASSCF, CASPT2, MRCI calculations as well as coupled-cluster (CC) calculations. All the results are illustrated in Fig.~\ref{Fig.H2o_MLDMRG}. It can be seen that the accuracy of ML-DMRG is gradually improved with the increasing \textit{M} value in the low-level subspace. If only 1 preserved state is used in low-level subspace, ML-DMRG can reproduce the result of CASCI in only the high-level subspace. It means that only the non-dynamical correlation in the high-level subspace is considered, while there is no dynamical correlation of low-level subspace being involved. When increasing the number of preserved states that used in low-level subspace, one would find that ground-state energy decreases gradually,
and finally the energy value (-76.3267187 Hartree, $M$=300 in low-level subspace) is even lower than that of CCSD(T) (-76.3265256 Hartree), and it is quite close to that of standard DMRG calculation using \textit{M}=3000 (difference is about 0.0005 Hartree).
It means that the electronic correlations within the low-level subspace and those between the high-level and low-level subspaces can be effectively incorporated upon increasing \textit{M} used in low-level subspace.

\begin{figure}[!htp]
 \begin{center}
   \includegraphics[scale=0.35, bb=0 75 1500 400]{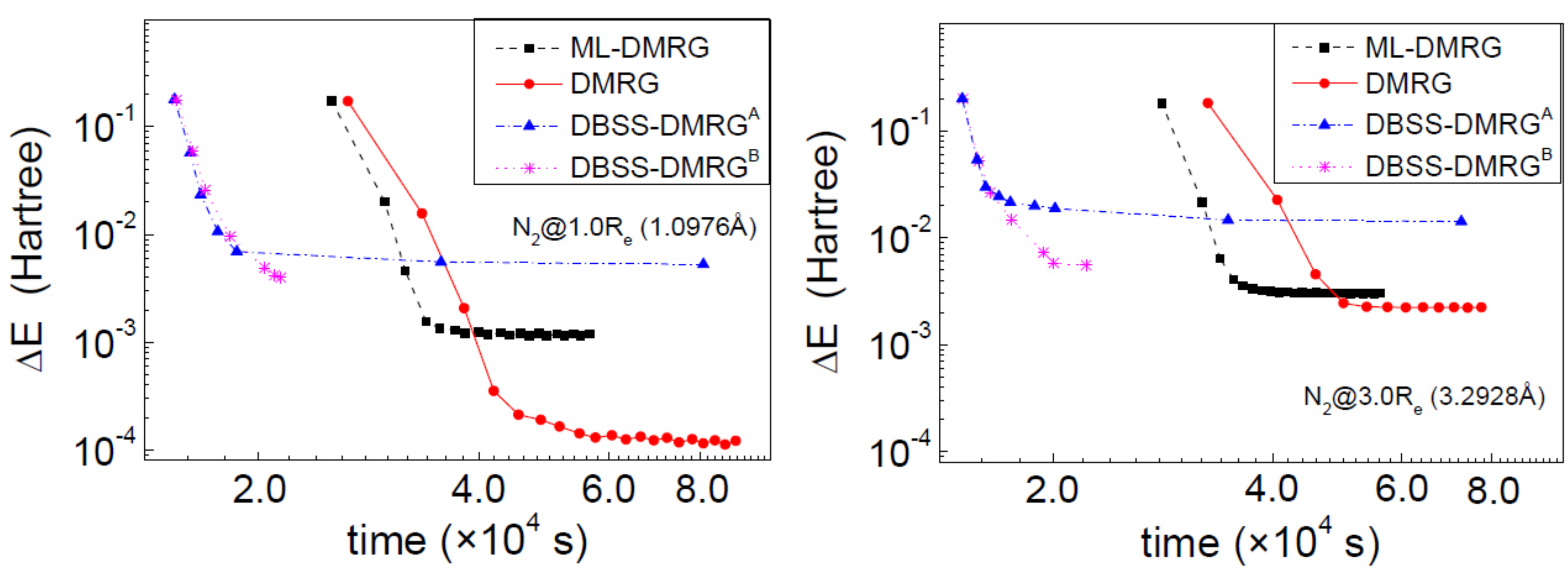}
 \end{center}
 \caption{Energy convergence of full orbital DMRG calculation of N$_2$ molecule at 1.0$R_e$ (1.0976 \AA, left part) and 3.0$R_e$ (3.2928 \AA, right part) intra-molecule distance.
The ML-DMRG calculations were implemented with $M$= 750, 3000, 1500, 750 in four different subspaces.
Every point of all the four methods represent the difference between the energy when sweeping on the middle of active orbitals and the final extrapolated energy. The latter parts of these lines are flattening out, showing that the calculations are converged at this level.
For comparisons, the DMRG calculations were implemented with $M$=3000 in the whole active space as well as two DBSS-DMRG calculations with different truncation cutoffs of the reduced density matrix $\rho_{tr}$ (A:1.0$\times10^{-7}$; B:1.0$\times10^{-8}$) are also shown.} All the calculations were implemented using 16-core Inspur clusters with Intel Xeon E5-2670 CPU.
\label{Fig.N2_MLDMRG}
\end{figure}

\quad Secondly, we extend the active space to all the orbitals of N$_2$ molecule with further partition of orbital levels, in order to reveal the capacity and behavior of ML-DMRG when dealing with large active spaces (around 60 orbitals). The N$_2$ molecule in both equilibrium geometry (1.0$R_e$, $R_e$=1.0976\AA) and near dissociation geometry (3.0$R_e$) with cc-pVTZ basis are used. The NOs calculated from CASPT2 are used as the basis when implementing DMRG or ML-DMRG calculations. In these calculations, the inner orbitals (\textit{1s}), valence orbitals (\textit{2s2p}), third-shell orbitals (\textit{3s3p3d}), and forth-shell orbitals (\textit{4s4p4d4f}) are the orbitals of the four different subspaces. When implementing ML-DMRG calculations, the numbers of preserved states in different subspaces are kept by the proportion of 0.25:1:0.5:0.25. It is clearly shown that, for moderate energy accuracy criterion (no less than 10$^{-3}$ Hartree) ML-DMRG calculations can have noticeable efficiency gain over standard DMRG calculations with fixed $M$ values (Fig.~\ref{Fig.N2_MLDMRG}). One may also notice that the DBSS strategy with $\rho_{tr}$=1.0$\times10^{-7}$ has significant computational speed advantages for low energy accuracy (around 10$^{-2}$ Hartree) calculations, which save more than 50\% computational time. However, more accurate DBSS-DMRG calculations with $\rho_{tr}$=1.0$\times10^{-8}$ can be easily out of the capability of current
computational resources and hardly give converged results because it usually requires too many states to be kept (more than 10000, see Fig. S1a \cite{Sx_ref}). Under such circumstances, using fixed $M$ value or the ML strategy will be a computational advantage for DMRG calculations. Therefore, one can get a useful hint: for achieving computational efficiency gains, the DBSS strategy provides a good choice for low-accuracy DMRG calculations, while the ML strategy can provide an alternative solution for medium-accuracy DMRG calculations.

\clearpage

\subsection{Accuracy of ML-DMRG-SCF}
\quad

\begin{figure}
 \begin{center}
   \includegraphics[scale=0.45, bb=0 75 1000 200]{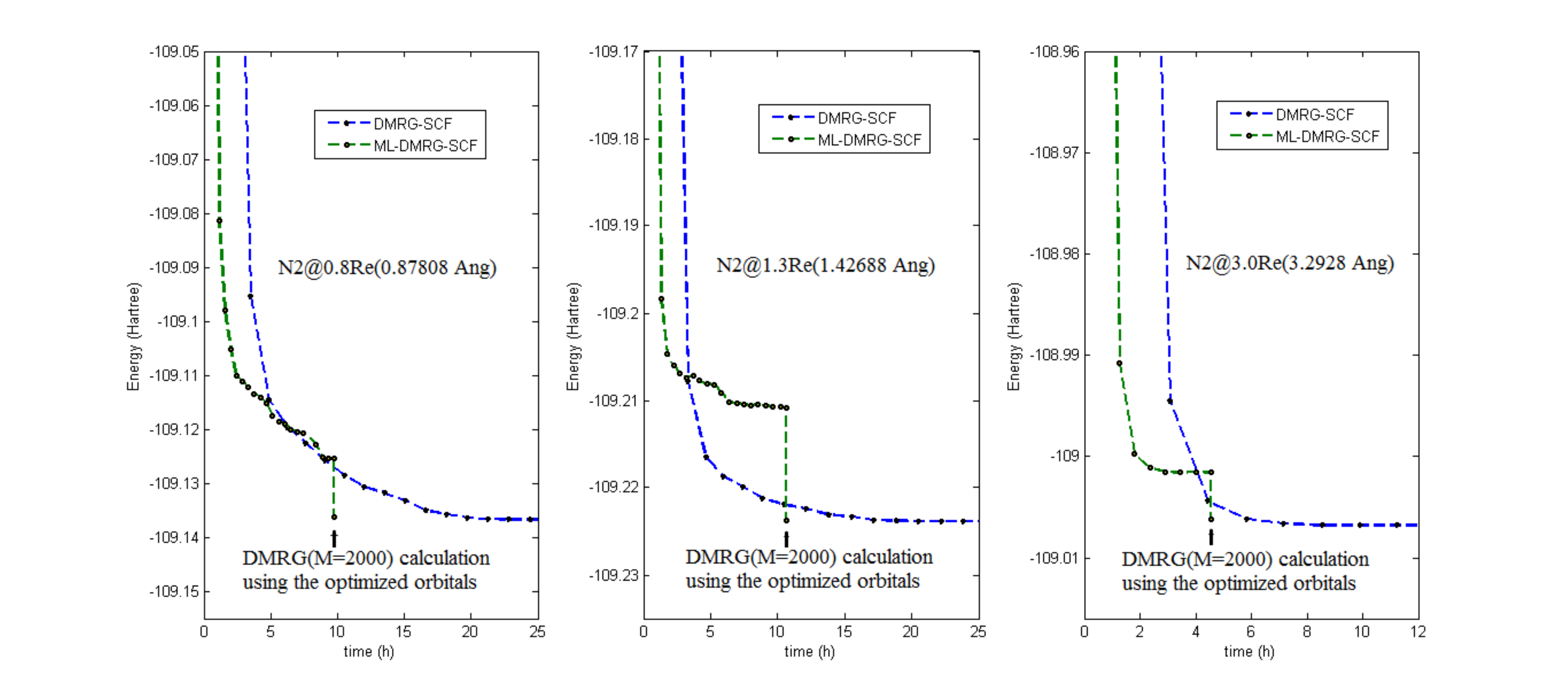}
 \end{center}
\caption{The convergence behavior of ML-DMRG-SCF (without orbital rotations between different active subspaces) and DMRG-SCF.}
\label{fig:frag_m}
\end{figure}

\quad We also build ML-DMRG-SCF
and apply it to the calculation of the N$_2$ molecule. Actually, (6e, 6o) should already give good descriptions of the dissociation process of N$_2$, if only a qualitative picture is required.
In ML-DMRG-SCF, the core (\textit{1s}) and valence orbitals (\textit{2s, 2p}) are chosen as the high-level orbitals with ($M$=2000), while the \textit{3s}, \textit{3p}, \textit{3d} orbitals are chosen as the low-level orbitals with ($M$=200). Here we implemented two types of orbital optimization strategies as shown in Fig.~\ref{Fig.ML-DMRG-SCF}: 1) Orbital rotations between different levels are allowed, i.e. original treatment or RASSCF-type treatment, and 2) these are not allowed, i.e. simplified treatment. The cc-pVQZ basis is used.
When the RASSCF-type rotations are turned on, we found that the ML-DMRG-SCF may diverge after several SCF iterations.
Once the inter-subspace rotations are turned off, the convergence behavior became much improved (Fig.~\ref{fig:frag_m}).
In the latter case, the results of ML-DMRG-SCF have stable deviations (10 $\sim$ 14 mH) when comparing with those by DMRG-SCF (Fig.~\ref{tab-N2pec}).
When the optimized orbitals from ML-DMRG-SCF are used as the basis to run further higher-level DMRG calculations, the deviations between results of DMRG and those of DMRG-SCF are less than 1 mH. It means the optimized orbitals from ML-DMRG-SCF calculation owns a good quality comparable to that by DMRG-SCF calculation. It also implies that the ML-DMRG-SCF could be used as the alternative to DMRG-CASSCF for yielding optimized orbitals, or be used as a pre-DMRG-SCF calculation in order to run a DMRG-SCF calculation more efficiently.

\begin{table}[!hbp]\centering
\begin{scriptsize}
 \begin{threeparttable}
 \caption{\label{tab-N2pec} Calculated energies (a.u.) of N$_2$ system in different geometry configurations by DMRG-SCF and related energy differences (a.u.) between
RASSCF$^a$, ML-DMRG-SCF$^{b,c}$, DMRG/ML-DMRG-SCF$^d$ results and them.}
  \begin{tabular}{cccccccccccccc}
  \hline
  \hline
 Method   & & 0.8$R_e$& 1.0$R_e$ & 1.3$R_e$ & 2.0$R_e$ &3.0$R_e$ &4.0$R_e$ \\
  \hline
 RASSCF$^a$        && 0.00824 & 0.00866 & 0.00239 & 0.00823 & 0.00849 & 0.00856 \\
 ML-DMRG-SCF$^b$   && 0.0068 & 0.0063 & 0.0090 & 0.0056 & 0.0049 & 0.0051 \\
 ML-DMRG-SCF$^c$   && 0.0112 & 0.0132 & 0.0138 & 0.0106 & 0.0101 & 0.0101 \\
 DMRG/ML-DMRG-SCF$^d$ && 0.00056 & 0.00097 & 0.00054 & 0.00008 & 0.00004 & 0.00004 \\
 DMRG-SCF  && -109.13671 & -109.36550 & -109.22427 & -109.01743  & -109.00660  & -109.00618 \\
  \hline
  \hline
  \end{tabular}
  \begin{tablenotes}
\item[a] Orbital rotations between different active subspaces are allowed.
Notice that extra attempts (e.g. start from pre-optimized orbital) may be used, since it may diverge during the ML-DMRG-SCF iterations in this case.
\item[b] No orbital rotations between different active subspaces.
\item[c] RASSCF(14e,28o) with CAS(6e,6o) as RAS2, two holes are allowed in RAS1 and two electrons are allowed in RAS3.
\item[d] DMRG(14e,28o) calculations ($M$=2000) with optimized orbitals from ML-DMRG-SCF$^2$.
  \end{tablenotes}
 \end{threeparttable}
\end{scriptsize}
\end{table}

\clearpage
\subsection{State-averaged ML-DMRG-SCF calculation of absorption spectra}

\begin{figure}[!htp]
 \begin{center}
   \includegraphics[scale=0.30, angle=180, bb=0 0 600 450]{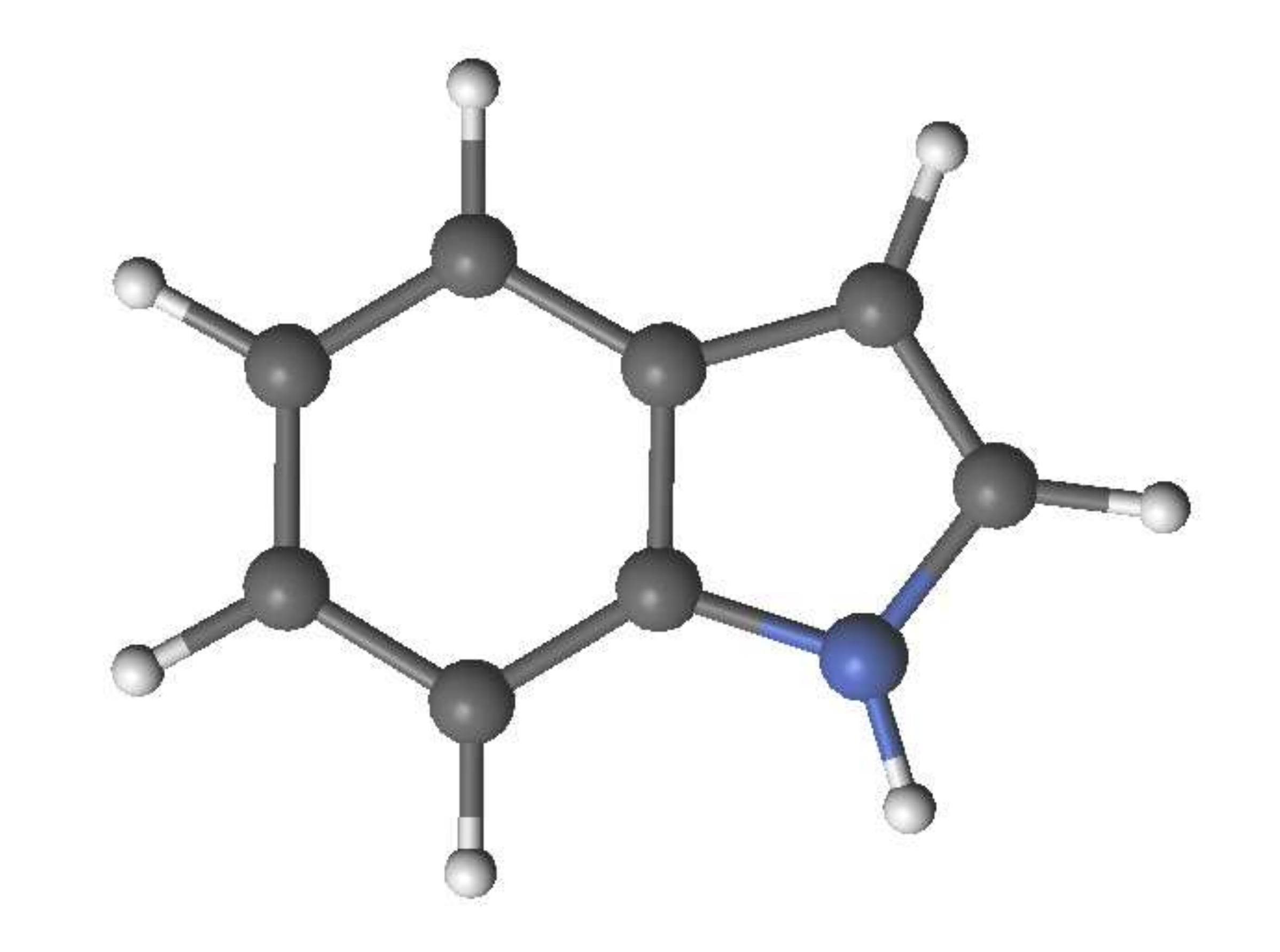}
 \end{center}
  \caption{Indole molecule, the geometry was optimized by BP86-D3/augTZ}
\label{Fig.indole}
\end{figure}

\quad Hereby, we implement the state-averaged ML-DMRG-SCF method in order to evaluate the electronic excitations in complicated molecular systems. As an example we choose to study the indole molecule (see Fig.~\ref{Fig.indole}). The indole chromophore is responsible for the low-energy absorption and emission spectra in order to understand the tryptophan spectroscopic properties \cite{indole_REV1, indole_REV2}, which are believed to be useful probes of local environment and dynamics in proteins \cite{indole1}. The studies on the electronic spectrum of indole have been almost exclusively focused on the position and nature of the two low-lying valence excited singlet states ($2^1A'$ and $3^1A'$), which were early labeled $^1L_b$ and $^1L_a$ states \cite{indole_early}. In a previous study, Serrano-Andr\'{e}s and Roos pointed out that a valence basis set and a valence active space do not guarantee that the obtained states have a pure valence character, and as such, the recommended procedure is therefore to always include diffuse functions in the basis set and Rydberg-type orbitals in the active space \cite{indole}.

\begin{figure}[!htp]
 \begin{center}
   \includegraphics[scale=0.35, bb=0   0  1300 700]{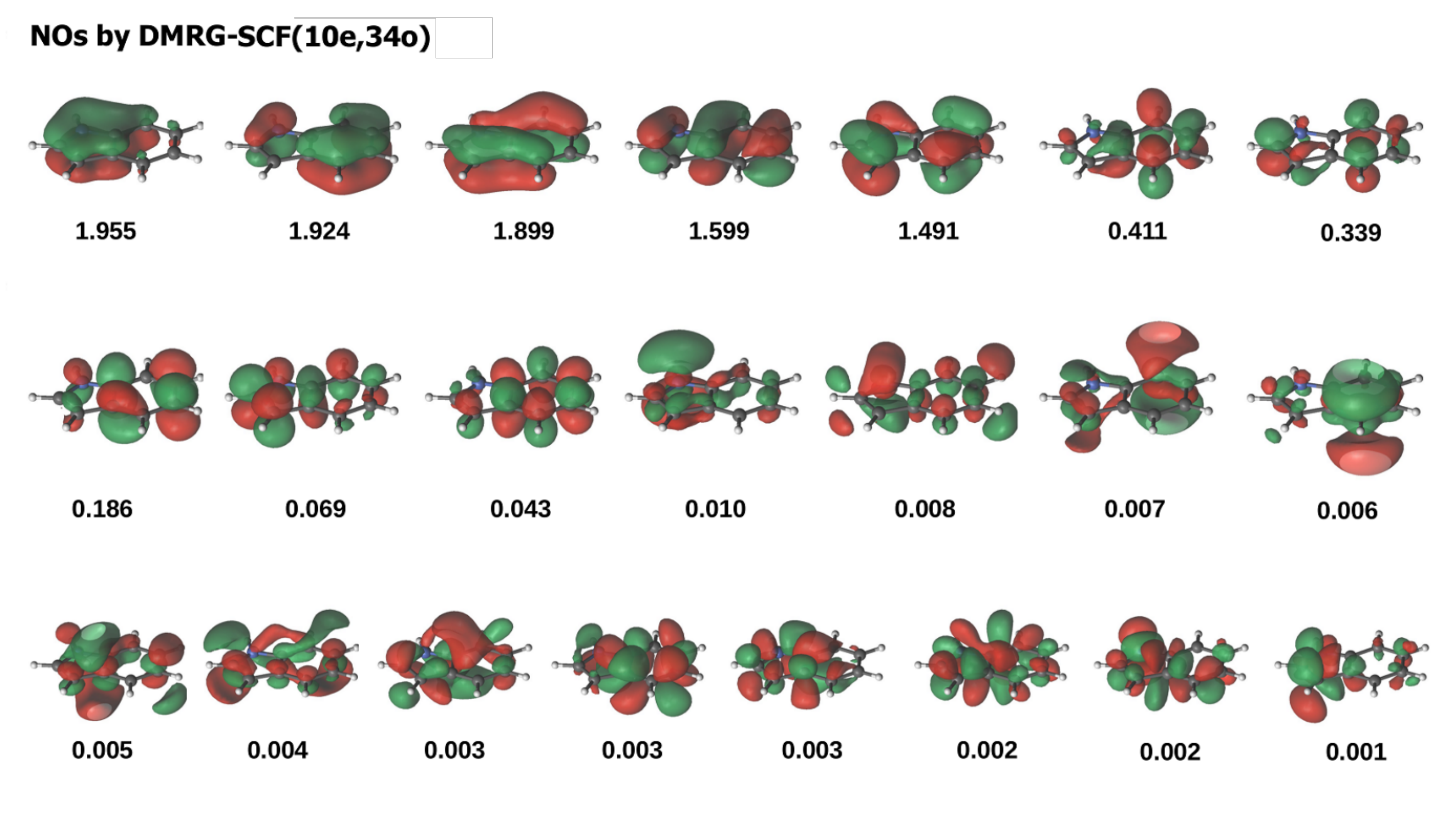}
   \includegraphics[scale=0.23, bb=0 -100 1750 50]{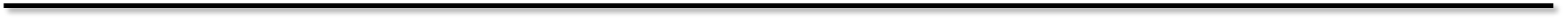}
   \includegraphics[scale=0.34, bb=0  100 1300 700]{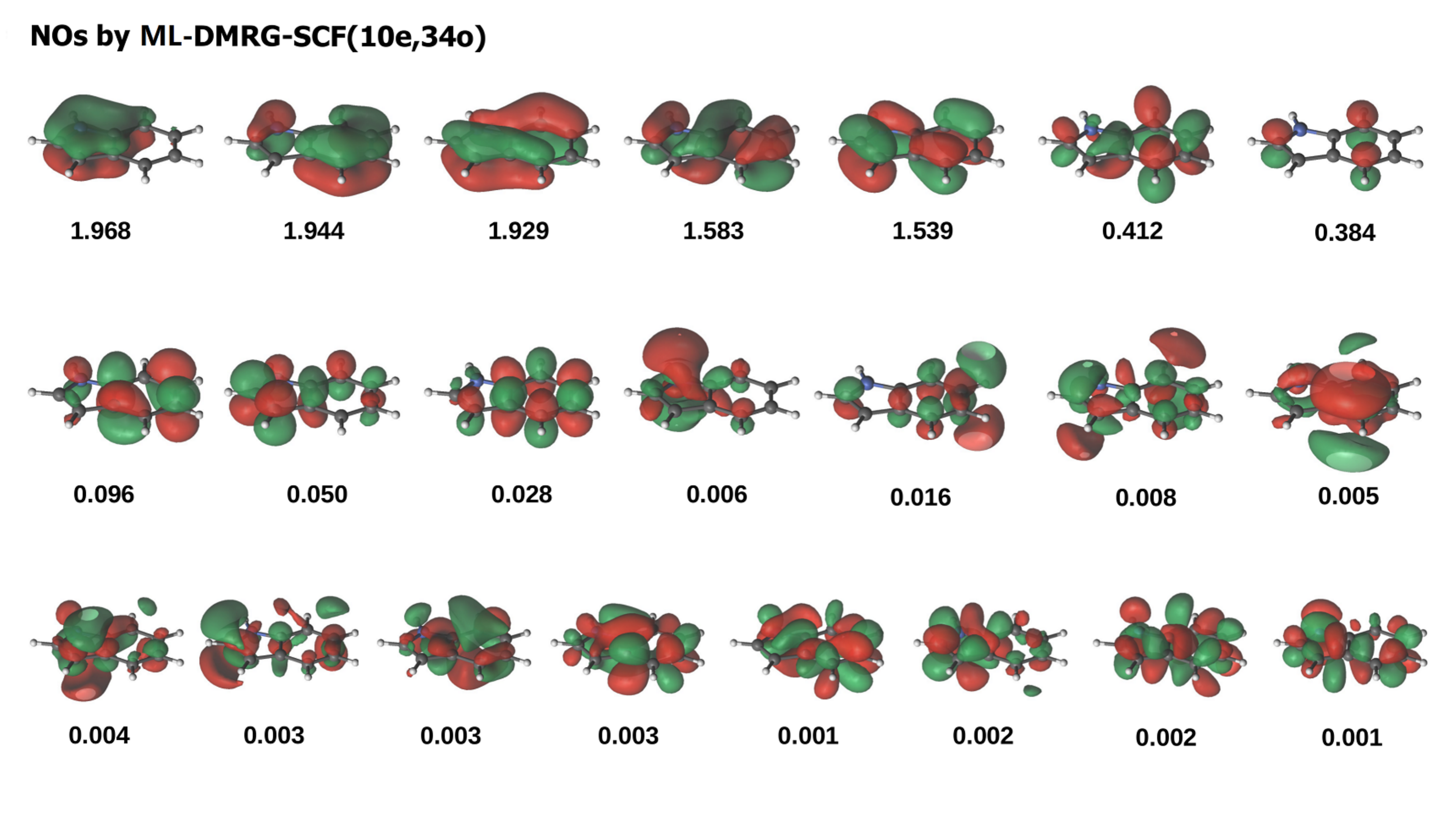}
 \end{center}
  \caption{The NOs as well as the NOONs of $a''$ orbitals in indole molecule. Top: Orbitals are optimized by state-averaged DMRG-SCF(10e,34o) with $M$=1000 (12$a'$+22$a''$ orbitals). Bottom: Orbitals are optimized by state-averaged ML-DMRG-SCF(10e,34o) with $M$=1000 in high-level subspace (6$a'$+12$a''$ orbitals), $M$=100 in low-level subspace (6$a'$+10$a''$ orbitals).}
\label{Fig.indole_ML_NOs}
\end{figure}

State-averaged ML-DMRG-SCF and state-averaged DMRG-SCF are used to calculate the excitation energies of indole molecule. When using ML-DMRG-SCF, the choice of active subspaces is as following: 1) valence orbitals (9 $\pi$) plus some Rydberg-type orbitals (6 $\sigma$ and 3 $\pi$) are treated as high-level orbitals with $M$=1000, while some of the other Rydberg-type orbitals (6 $\sigma$ and 10 $\pi$) are treated as low-level orbitals with $M$=100. The results are listed in Table.~\ref{tab-indole}.
It is shown that state-averaged ML-DMRG-SCF could give qualitatively correct and quantitatively reasonable descriptions of $2^1A'$ and $3^1A'$ states. The shift between these two states is determined to be 0.69 eV by ML-DMRG-SCF, close to that by the state-averaged DMRG-SCF calculation (0.75 eV). When the further DMRG calculation is implemented with the optimized orbitals from state-averaged ML-DMRG-SCF, we can get the results nearly the same as that by state-averaged DMRG-SCF. Additionally, we illustrate the calculated NOs of $a''$ symmetry in Fig.~\ref{Fig.indole_ML_NOs} ($2^1A'$, $3^1A'$ states are belong to $\pi - \pi^*$ type excitation, thus only $a''$ orbitals are involved). It can be seen that the NOs obtained from state-averaged ML-DMRG-SCF and those from state-averaged DMRG-SCF are similar with each other, whatever in the shapes or in the natural orbital occupation numbers (NOONs).

\begin{table}[!hbp]\centering
\begin{footnotesize}
 \begin{threeparttable}
 \caption{\label{tab-indole} Excitation energies (in eV) of $2^1A'$ and $3^1A'$ states (by state-averaged calculation for 5 states) in indole system and the energy shift values (in eV) between them. The geometry was optimized by BP86-D3/augTZ and the aug-cc-pVTZ basis set is used when implementing DMRG calculations.}
  \begin{tabular}{cccccccccc}
  \hline
  \hline
    & & \multirow{2}{*}{CASSCF$^a$} & DMRG- & ML-DMRG- &  DMRG / ML-&  DMRG- & \multirow{2}{*}{exptl.$^f$}\\
    & &   & SCF$^b$ & SCF$^c$ &  DMRG-SCF$^d$ &  SCF$^e$ &  \\
  \hline
 $2^1A'$  &&  4.67 & 4.78 & 4.87 & 4.79 & 4.76 & 4.37 \\
 $3^1A'$  &&  5.60 & 5.58 & 5.56 & 5.53 & 5.51 & 4.77 \\
 $\Delta$ &&  0.93 & 0.80 & 0.69 & 0.74 & 0.75 & 0.40 \\
  \hline
  \hline
  \end{tabular}
  \begin{tablenotes}
\item[a] CASSCF(10e,12o), $0a'$ and $12a''$ orbitals are chosen.
\item[b] DMRG-SCF(10e,18o) with $M$=1000, $6a'$ and $12a''$ orbitals are chosen.
\item[c] ML-DMRG-SCF(10e,34o), $6a'$ and $12a''$ orbitals are chosen as the high-level orbitals with $M$=1000, the other $6a'$ and $10a''$ orbitals are chosen as the low-level orbitals with $M$=100. The orbital rotations between different levels are not allowed. Typical time for per-SCF iteration is around 5h using HP ProLiant DL380p clusters (CPU : Xeon E5-2670$\times$2, 16 cores).
\item[d] DMRG(10e,34o) calculation ($M$=1000) using optimized orbitals from c.
\item[e] DMRG-SCF(10e,34o) calculation ($M$=1000). Typical time for per-SCF iteration is around 11h using the same clusters as [c].
\item[f] ref.~\cite{indole}.
  \end{tablenotes}
 \end{threeparttable}
\end{footnotesize}
\end{table}

\clearpage
\subsection{Orbital ordering effect and entanglement analysis}

\quad It is well-known that the orbital ordering would significantly affect the DMRG performances, and there were many informative suggestions \cite{DBSS, Ma13, DMRGRCM, Block, Quan_inf, DMRG_env, DMRG_ctrl, DMRG_qinf, Rissler06, Legeza03, Legeza03-1} on how to construct an optimal orbital ordering before implementing the computationally expensive DMRG calculations. When using canonical-type orbitals, one reasonable choice is to use the bonding/anti-bonding ordering, which has been recommended by many groups \cite{DBSS, Ma13, DMRGRCM}. However, in most cases it is not straightforward and it is also extremely time-consuming to pick up the desired bonding/anti-bonding ordering manually, and as such, there are some other suggested automatic strategies, such as reverse Cuthill-Mckee algorithm \cite{RCM, Legeza03-1, DMRGRCM}, genetic algorithm \cite{Moritz05}, Minimun Bandwidth by Perimeter Search \cite{Mizukami13}, and algorithms from graph theory \cite{Graph, Block}. Since the ML-DMRG strategy changes the orbital ordering when partitioning orbital spaces, it is of great importance to also investigate the ML-type orbital ordering effect on DMRG calculations.

For illustrating the entanglement behavior and orbital ordering effect within ML-DMRG calculations, here we show our benchmark calculations of Cr$_2$ system at two geometrical configurations (inter-atomic distance 1.5\AA \ and 2.8 \AA). In our calculations cc-pVTZ basis is used with a 12e-in-42o active space. The NOs from CASPT2(12e,12o) are used and the partition of active spaces is as following: the valence orbitals (12 orbitals, i.e. \textit{3d4s} orbitals) are chosen as the high-level orbitals; 16 orbitals which is based on NOONs (mainly composed by \textit{4p4d} orbitals) are chosen as medium-level orbitals; other 14 orbitals, which are also based on NOONs, are chosen as the low-level orbitals. Different Davidson convergence thresholds ($10^{-4}$, $10^{-5}$, $10^{-6}$) are used for parallel comparisons.

\begin{figure}[!htp]
 \begin{center}
   \includegraphics[scale=0.25,bb=0 50 1600 1200]{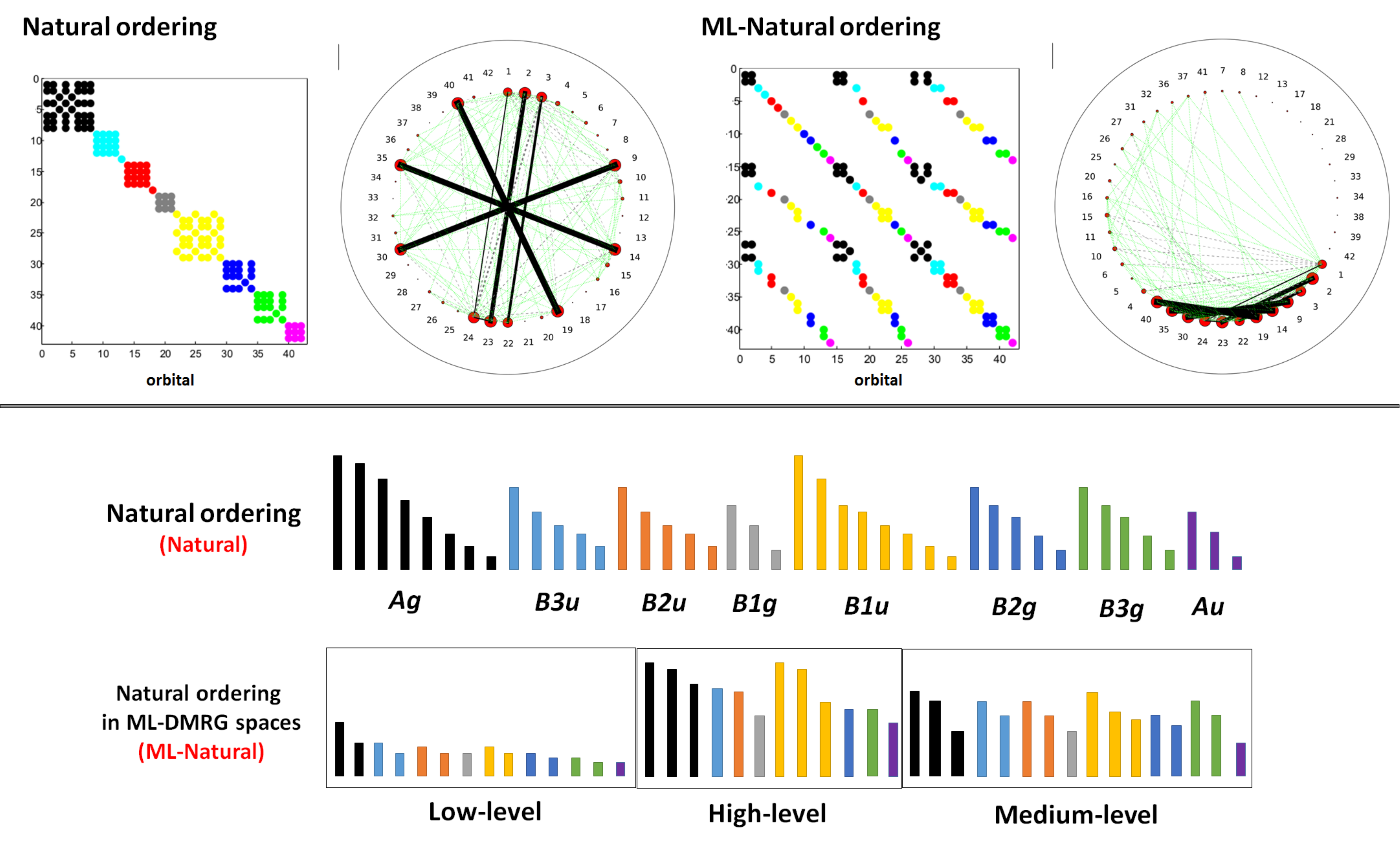}
 \end{center}
  \caption{Schematic illustration of natural ordering and its ML expansion in Cr$_2$ ($r$=2.8 \AA) systems with 12e-in-42o. Here one column represents one orbital and the height of the column reflects the NOON. Their one-body integrals (shown is squares) and the entanglements [1'-orbital entropy and mutual information (i.e. orbital interaction), shown in circles] are also illustrated. In the entanglements circles, the magnitude of red point reflects the single orbital entropy; the lines that link two orbitals represent the mutual information. The black solid line denotes the mutual information larger than 0.1, gray dashed line for 0.01, and green line for 0.001. The entanglements are calculated by {\sc Maquis} (DMRG) using \textit{M}=2000 states.}
\label{Fig.ordering}
\end{figure}

\begin{figure}[!htp]
 \begin{center}
   \includegraphics[scale=0.25,bb=0 0 1600 1050]{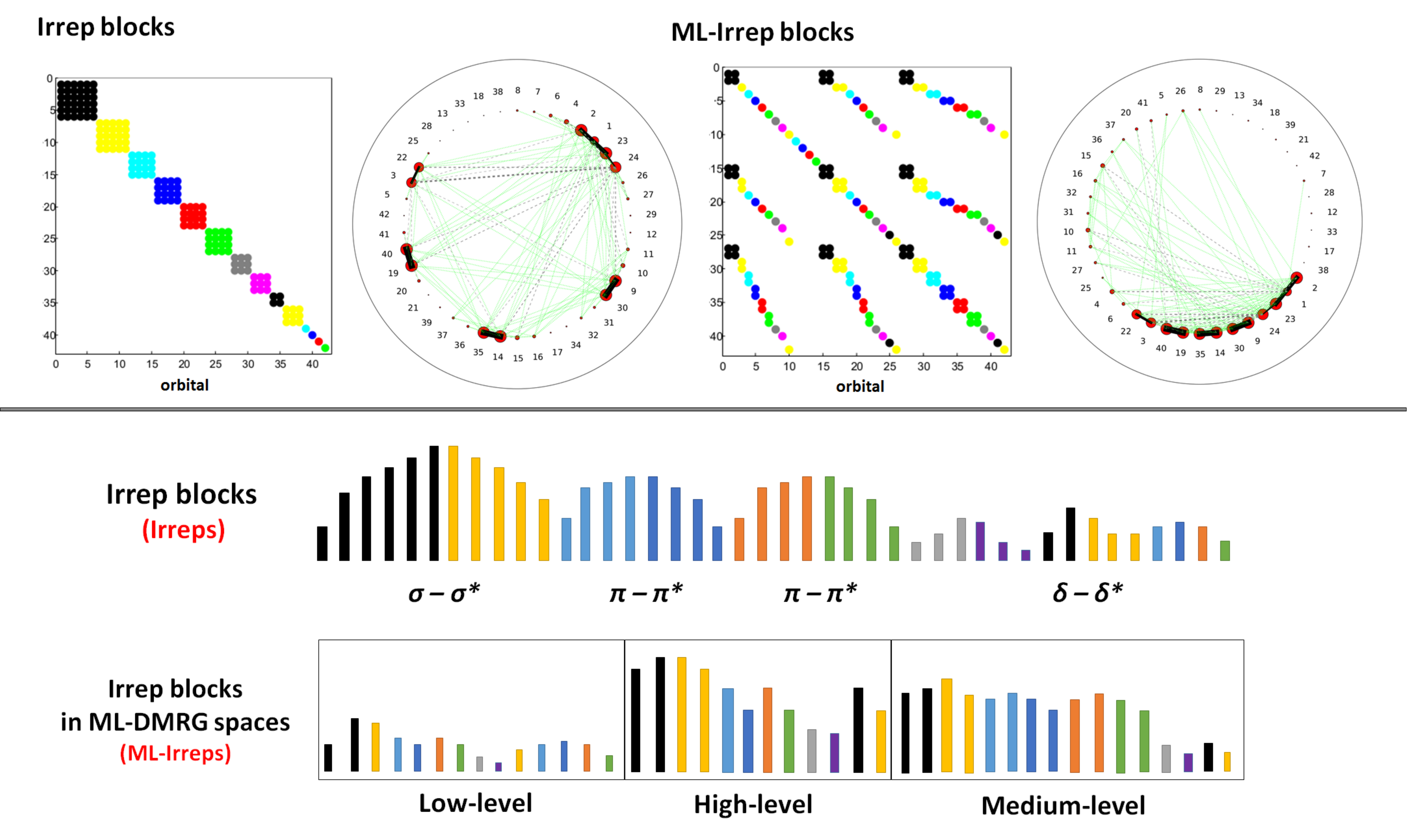}
   \includegraphics[scale=0.25,bb=0 50 1600 1050]{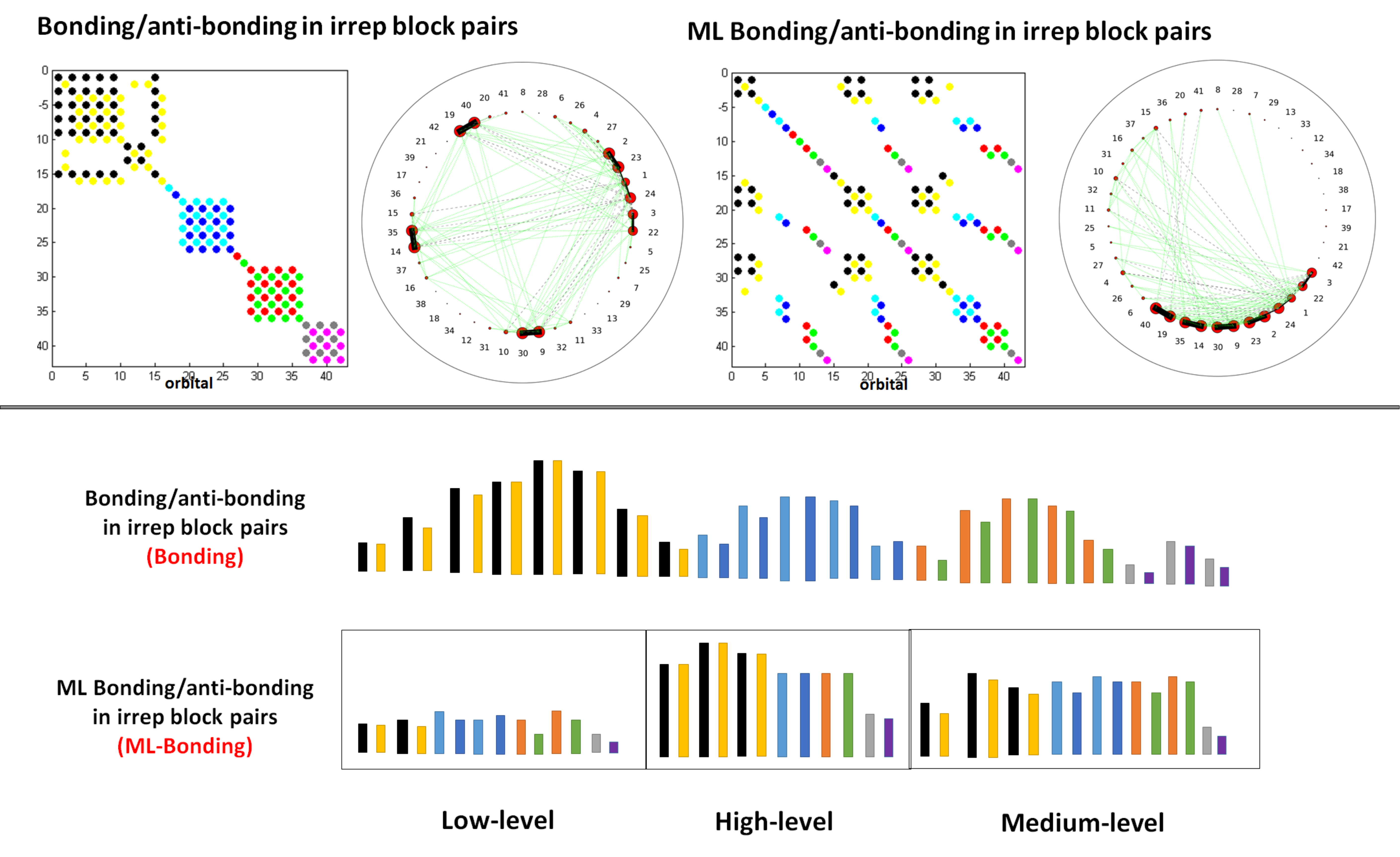}
 \end{center}
  \caption{Same as Fig.~\ref{Fig.ordering}, schematic illustration of  irrep blocks ordering, bonding/anti-bonding ordering in irrep block pairs, and their ML expansions in Cr$_2$ ($r$=2.8 \AA) system with 12e-in-42o.}
\label{Fig.ordering2}
\end{figure}

\begin{figure}[!htp]
 \begin{center}
   \includegraphics[scale=0.525, bb=0 50 900 575]{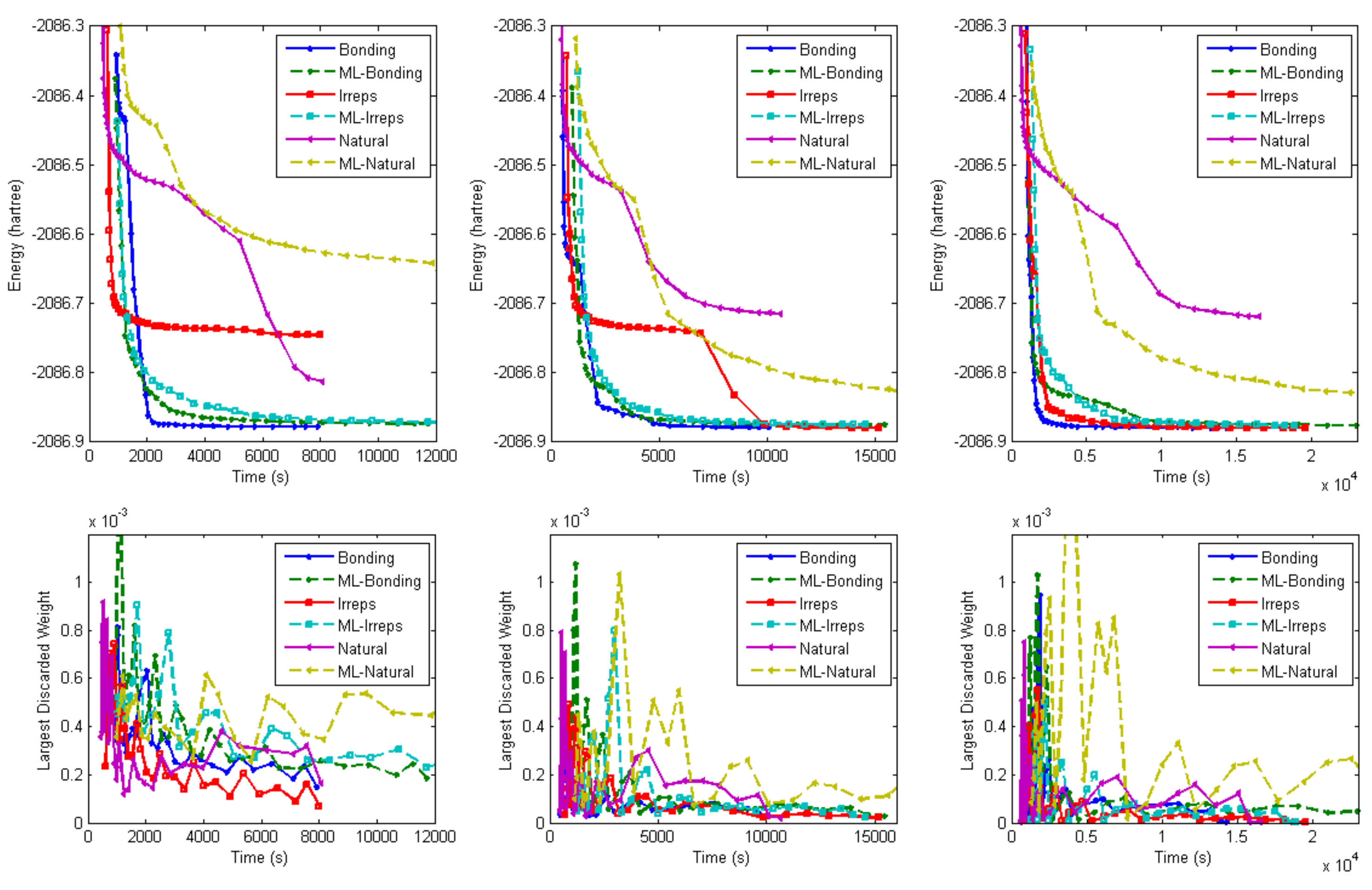}
 \end{center}
  \caption{The convergence behaviors of DMRG calculations of Cr$_2$ with 2.8 \AA\ inter-atomic distance using different orbital orderings and different convergence thresholds.
The Davidson threshold values are $10^{-4}$, $10^{-5}$, $10^{-6}$ from left to the right. The \textit{M} values were increased regularly: 50, 100, 200, 400, 800, 1200, 1600, and 2000. There are 4 sweeps for each \textit{M} value. The largest discarded weights during sweeps are also listed at the bottom. HP ProLiant DL380p clusters (CPU: Xeon E5-2670$\times$2, 16 cores) are used in calculations.}
\label{Fig.convergence2}
\end{figure}

Fig.~\ref{Fig.ordering} and Fig.~\ref{Fig.ordering2} show the orbital ordering schemes and their partition of 1-body integrals as well as the entanglements (1'-orbital entropy and mutual information).
Two facts can be easily found for the ML-DMRG orbital reordering strategy compared with the original ones: 1) the one-body integrals would no longer locate at the thick diagonal part but they will be split into several thin sub-bands; 2) the entanglements, both of the single-orbital entropy and mutual information, are no long well distributed in the whole orbital space, and they are concentrated in the high-level subspace.
This may bring an advantage of quicker DMRG convergence in each subspace since the major entanglements are more localized than before. But it may also bring a disadvantage of slower DMRG convergence when accounting the unbalanced inter-subspace entanglements \cite{Legeza03}.
The comparison of DMRG convergence behaviors is illustrated in Fig.~\ref{Fig.convergence2} for 2.8 \AA \ Cr-Cr distance and those for 1.5 \AA \ Cr-Cr distance in Fig. S2 \cite{Sx_ref}.
Interestingly, when loose or moderate Davidson threshold ($10^{-4}$ or $10^{-5}$) for wavefunction accuracy was used in DMRG calculations, the ML-type ordering may offer easier convergences to the lower energy than their original counterparts in the beginning, especially when the original orderings are well but not perfectly reordered (e.g. the irrep blocks ordering in our calculations). This can be ascribed to the above mentioned advantage of ML-type orderings in accounting intra-band interactions for moderately-accurate wavefunction calculations and also its shortcoming in describing the inter-band interactions. This provides a useful hint for future DMRG calculations: in the case of moderate accuracy requirement of the wavefunction, ML-type orbital ordering can be used as an efficient ``warm-up" strategy to save substantial computational costs.

\begin{figure}[!htp]
 \begin{center}
   \includegraphics[scale=0.525, bb=0 50 1300 800]{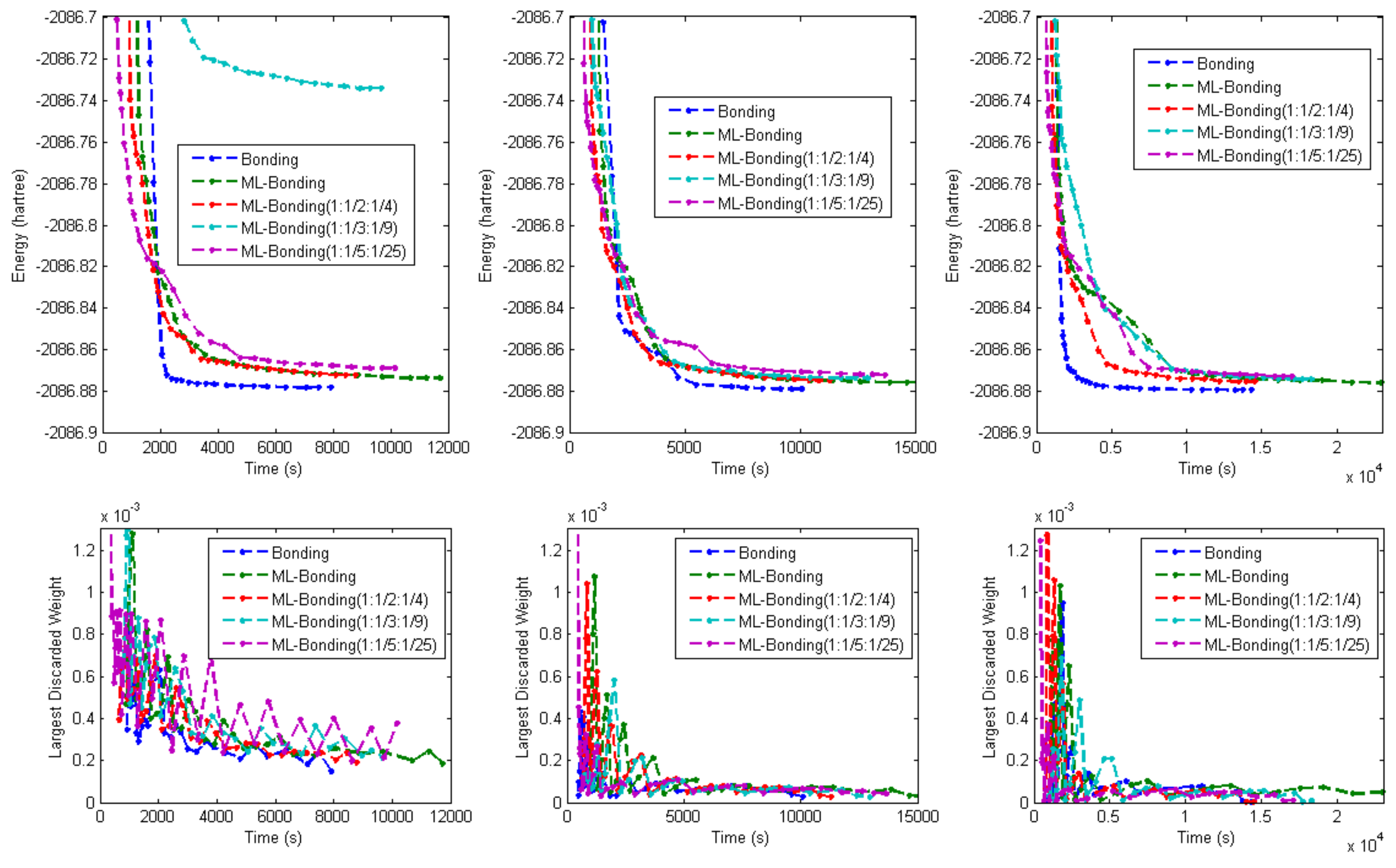}
 \end{center}
  \caption{The convergence behaviors of DMRG calculations of Cr$_2$ with 2.8 \AA\ inter-atomic distance using ratio in active subspaces and different convergence thresholds. The calculating condition is as the same as that shown in Fig.~\ref{Fig.convergence2} with the bonding/anti-bonding ordering in irrep block pairs. Notice that only the energy region from -2086.7 to -2086.9 (in hartree) are shown for better presentations.
}
\label{Fig.convergence22}
\end{figure}

When we further examine the efficiency difference between ML-DMRG calculations and DMRG calculations with fixed \textit{M} values within the same fixed ML-bonding/anti-bonding orbital orderings, one can also find from the top panel of Fig.~\ref{Fig.convergence22} that ML-DMRG calculations (with different ratio of preserved states shown in legend) do not have significant superiority over their counterparts. This can be understood from the discarded weight evolution pictures in the bottom panel of Fig.~\ref{Fig.convergence22}, where the discarded weights in ML-DMRG calculations are generally much larger than DMRG calculations with fixed \textit{M} values (blue line) in the case of the complicated Cr$_2$ system which is much different from the earlier N$_2$ system (see Fig. S1b \cite{Sx_ref}). This implies that ML-DMRG calculations may be not efficient for very complicated correlation systems because such cases usually requires a large number of \textit{M} at each orbital,
whereas ML-type orderings can still be suggested for DMRG calculations with fixed \textit{M} values for such complicated cases according to our above discussion.

\clearpage
\section{Conclusion}
\label{sec:conclusion}
\quad In this paper, we build the DMRG algorithms in the ML-type framework and investigate their efficiency and accuracy performances compared with their standard DMRG counterparts through computational examples of H$_2$O, N$_2$, Cr$_2$ and indole. Our major
findings can be outlined as follows:

\begin{enumerate}
  \item Chemical intuition-based multi-level control of the large active space can be combined with DMRG calculations. A noticeable computational efficiency gain with a little price of energy inaccuracy (normally few mHartrees) can be obtained by ML-DMRG calculations in contrast to those standard DMRG calculations with fixed $M$ values.
  \item The optimized orbitals obtained from ML-DMRG-SCF calculations are similar to those from the DMRG-CASSCF calculations. Comparing to RASSCF, a larger high-level active subspace can be used in ML-DMRG-SCF.
  \item ML-type hierarchical orbital ordering could offer increased computational efficiency for DMRG calculations with medium accurate Davidson threshold, since the major intra-subspace entanglements are more localized within such a new ordering. However, it will lose computational advantage for highly accurate Davidson threshold due to its inefficiency in describing the delocalized inter-subspace entanglements.
\end{enumerate}

The above findings concerning on the energetic accuracy and computational efficiency as well as the energy convergence behavior of ML-DMRG type calculations provide useful
information for constructing more economic protocols for future large-scale DMRG calculations. For example,
ML-DMRG-SCF could be taken as a pre-DMRG-CASSCF calculation in order to run a DMRG-CASSCF calculation more efficiently.
The hierarchical ML-bonding/anti-bonding orbital ordering may also be recommended for the DMRG calculations in large and complicated active spaces.

\section*{Acknowledgments}

\quad This work was supported by the National Natural Science Foundation of China (Grant No. 21373109).
We thank M. Reiher for stimulating discussions and providing us the development version of the MOLCAS package to perform some of the numerical calculations in this paper. We are also grateful to S. Knecht, S. Keller, C. Liu, and W. Hu for many helpful conversations.

\end{document}